# DESIGN AND PERFORMANCE ANALYSIS OF HYBRID ADDERS FOR HIGH SPEED ARITHMETIC CIRCUIT

Rajkumar Sarma[1] and Veerati Raju[2]

[1]School of Electronics Engineering, Lovely Professional University, Punjab (India)
rajkumar.sarma86@gmail.com
[2]Department of VLSI, Lovely Professional University, Punjab (India)
rajureddyv@gmail.com

## ABSTRACT

*Adder cells using Gate Diffusion Technique (GDI) & PTL-GDI technique are described in this paper. GDI technique allows reducing power consumption, propagation delay and low PDP (power delay product) whereas Pass Transistor Logic (PTL) reduces the count of transistors used to make different logic gates, by eliminating redundant transistors. Performance comparison with various Hybrid Adder is been presented. In this paper, we propose two new designs based on GDI & PTL techniques, which is found to be much more power efficient in comparison with existing design technique. Only 10 transistors are used to implement the SUM & CARRY function for both the designs. The SUM and CARRY cell are implemented in a cascaded way i.e. firstly the XOR cell is implemented and then using XOR as input SUM as well as CARRY cell is implemented. For Proposed GDI adder the SUM as well as CARRY cell is designed using GDI technique. On the other hand in Proposed PTL-GDI adder the SUM cell is constructed using PTL technique and the CARRY cell is designed using GDI technique. The advantages of both the designs are discussed. The significance of these designs is substantiated by the simulation results obtained from Cadence Virtuoso 180nm environment.*

## 1. INTRODUCTION

Addition is one of the fundamental arithmetic operations. It is used extensively in many VLSI systems such as application specific DSP architectures and microprocessors. In addition to its main task, which is adding two binary numbers, it is the nucleus of many other useful operations such as subtraction, multiplication, division, address calculation, etc. In most of these systems the adder is part of the critical path that determines the overall performance of the system. That is why enhancing the performance of the 1-bit full-adder cell (the building block of the binary adder) is a significant goal. Recently, building low-power VLSI systems has emerged as highly in demand because of the fast growing technologies in mobile communication and computation. The battery technology doesn't advance at the same rate as the microelectronics technology. There is a limited amount of power available for the mobile systems. So designers are faced with more constraints: high speed, high throughput, small silicon area, and at the same time, low-power consumption. So building low-power, high-performance adder cells is of great interest. Figure 1 shows the power consumption breakdown in a modern day high-performance microprocessor. The data path consumes roughly 30% of the total power of the system. Adders are an extensively used component in data paths and, therefore, careful design and analysis is required for these units to obtain optimum performance. On the other hand, as discussed in [4], we can see from the Figure 1 that clock signals consumes 45% of the total power, which is very high in fact. As power dissipation has become one of the most important constraints in the design flow of modern processors, therefore, under this common scenario, it has become extremely important to consider the power consumption of any proposed module when there are non-transitioning input data or there is no clock signal activity.

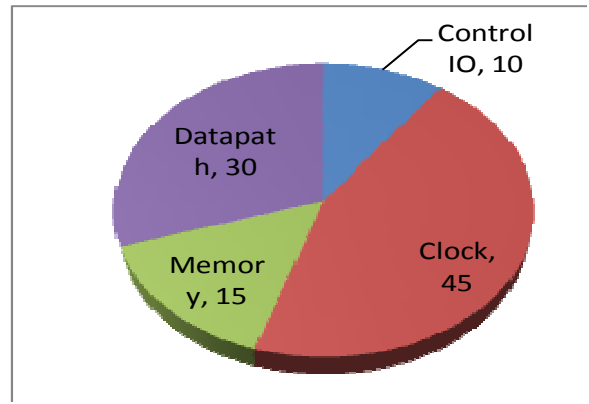

Figure 1. Shows the power consumption breakdown in a modern day high-performance microprocessor

Very often, the utmost integrated circuit performances are restricted by how best the arithmetic operators are implemented in the cell library provided to the designer for the synthesis. As the complexity of arithmetic circuits grows with increasing processor bus width, energy consumption is becoming more important now than ever due to the increase in the number and density of transistors on chip and faster clock. Different CMOS logic styles have evolved for the development of cell libraries. They are likely to perpetuate the ability to further reduce the cost-per-function and improve the performance of integrated circuits. With the lowering of threshold voltage in ultra deep submicron technology, lowering the supply voltage appears to be the most eminent means to reduce power consumption. However, lowering supply voltage also increases circuit delay and degrades the drivability of cells designed with certain logic styles. For example the goal to extend the battery life span of portable electronics is to reduce the energy expended per arithmetic operation, but low-power consumption need not necessarily implies low energy. To execute an arithmetic operation, a circuit can consume very low power by clocking at extremely low frequency but it may take a very long time to complete the operation.

Several logic styles have been used in the past to design full adder cells. Each design style has its own merits and demerits. Classical designs of full adders normally use only one logic style for the whole full-adder design. For example the static CMOS design, Pass Transistor Logic (PTL), Transmission Gate Logic and Complementary Pass Transistor Logic (CPL) etc. There are various advantages of one with another (i.e. if we compare different logic styles), which is discussed in the literature [2], [3], [4] & [5].

In this paper a full Adder cell using PTL (Pass transistor logic) & GDI (gate Diffusion input) are suggested. GDI technique allows reducing power consumption, propagation delay and low PDP (power delay product) as well. Whereas Pass Transistor Logic (PTL) reduces the count of transistors used to make different logic gates, by eliminating redundant transistors. In the next Section an introduction to GDI design is described. In Section III, IV & V different traditional logic styles, different hybrid adders & Proposed Adder is been discussed. The comparison of different adder with the proposed adder is been done in Section VI. Finally the advantages of Proposed GDI design and Proposed PTL-GDI design are discussed in the conclusion section.

## 1.1 Introductions to GDI (Gate Diffusion Input)

GDI method is based on the use of a simple cell as shown in figure 2. At the first look the design is seems to be like an inverter, but the main differences are 1) GDI consist of three inputs- G (gate input to NMOS/PMOS), P (input to source of PMOS) and N (input to source of NMOS). (2) Bulks of both NMOS and PMOS are connected to N or P (respectively), so it can be arbitrarily biased at contrast with CMOS inverter. Figure 4 shows the basic GDI cell.

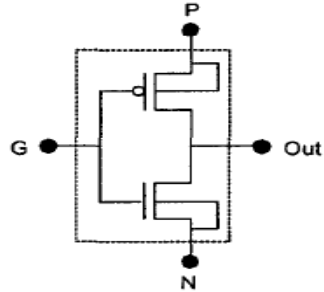

Figure 2. Basic GDI cell

Table 1 shows, as discussed in [9], how a simple change of the input configuration of the simple GDI cell corresponds to very different Boolean functions. Most of these functions are complex (6- 12 transistors) in CMOS, as well as in standard PTL implementations, but very simple (only 2 transistors per function) in GDI design method.

Table 1: Various logic functions of GDI cell for different input configurations

| N | P | G | Out | Function |
|---|---|---|-----|----------|
| '0' | B | A | $\bar{A}$B | F1 |
| B | '1' | A | $\bar{A}$+B | F2 |
| '1' | B | A | A+B | OR |
| B | '0' | A | AB | AND |
| C | B | A | $\bar{A}$B+AC | MUX |

## 1.2 Introductions to PTL (Pass Transistor Logic)

Another very popular design is Pass Transistor Logic design. When an NMOS or PMOS is used alone as an imperfect switch, we sometimes call it a Pass Transistor. PTL reduces the numbers of transistors used to make different logic gates, by eliminating excess amount of transistor. Transistors are used here as switches to pass logic levels between nodes of a circuit, instead of as switches connected directly to supply voltages (Vdd).

## 2. DIFFERENT HYBRID ADDERS

Several low-power and high-performance 1-bit hybrid Full Adder cells had been reported in the literature [1]. Here four different hybrid Full Adder Cells, which were reported to have better performance than others are reviewed and analyzed. The adders considered in this work were designed using traditional implementing methods, i.e. they use only transistors and no input capacitors are used.

In Radhakrishnan Adder [5] a minimal transistor CMOS pass network XOR-XNOR cell that is fully compensated for threshold voltage drop in MOS transistors, is presented by the author. This new cell can reliably operate within certain bounds when the power supply voltage is reduced to certain level. It uses only six transistors for the combined XOR-XNOR cell and can operate reliably when the supply voltage is scaled down, as long as the voltage is not allowed to fall below double of threshold voltage. The total number of transistor used here for full adder operation is 14. The Design circuit is shown in figure 3.

The Chang Adder [3] uses 26 transistors and it utilizes a modified low-power XOR/XNOR circuit. In this circuit worst case delay problems due to logic transitions are solved by adding more transistors; however, these additional transistors increase the power consumption of the full adder cell. The Design circuit is shown in figure 4.

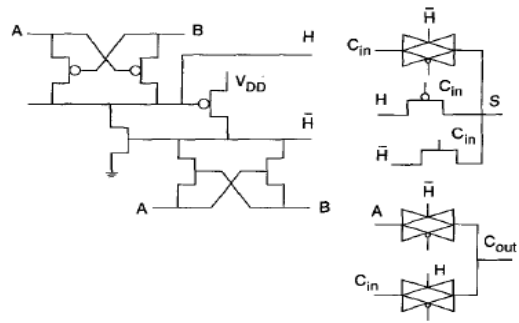

Figure 3. Radhakrishnan Adder [5]

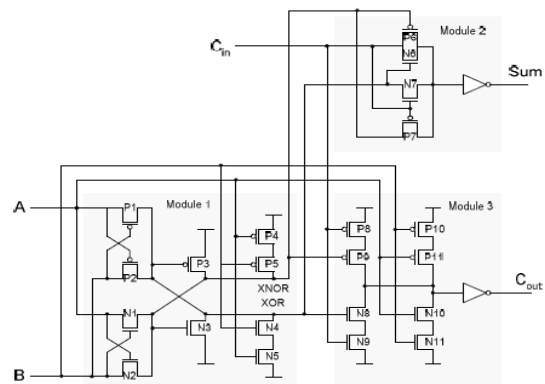

Figure 4. Chang Adder [3]

The Goel Adder [4] uses a XOR–XNOR circuit which can produce balanced full swing output. It has high-speed operation due to the cross-coupled PMOS Pull-up transistors providing the intermediate signals quickly and a hybrid- MOS output stage with a static inverter at the output. The Design circuit is shown in figure 5.

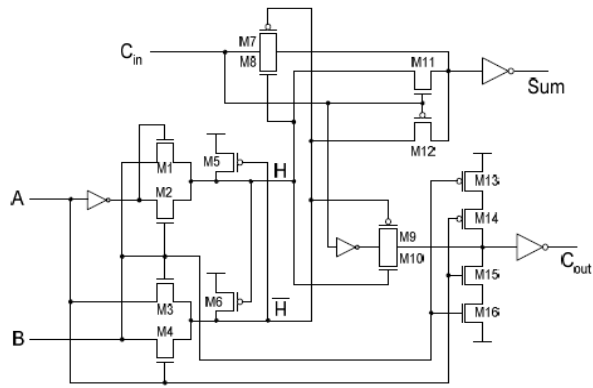

Figure 5. Goel Adder [4]

The Agarwal Adder [2] uses the Complementary Pass transistor Logic (CPL). This adder is mainly composed by NMOS transistors with pull–up PMOS transistors to obtain full swing output voltage. Due to positive feedback and use of NMOS transistors, the circuit is inherently fast. This adder has a balanced structure with respect to generation of SUM and CARRY signals. The Design circuit is shown in figure 6.

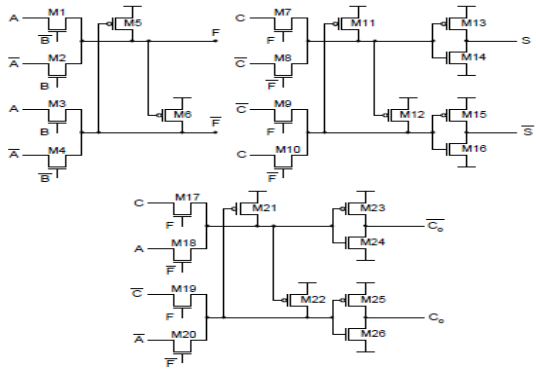

Figure 6. Agarwal Adder [2]

## 3. PROPOSED ADDERS

The designs are based on XOR-XOR based full adder style. The basic SUM as well as CARRY functionality is as follows:

$$H = A\ XOR\ B$$
$$SUM = H\ XOR\ Cin$$
$$CARRY = H'A + HCin$$

### 3.1 Proposed PTL-GDI Adder

In this design the SUM cell is designed using Pass Transistor Logic (PTL) and the CARRY cell is designed using Gate Diffusion Technique (GDI). Firstly the H function is been generated using two PMOS and two NMOS transistors and then using Cin and H as input the SUM function is obtained. The total number of transistor used is eight to obtain the SUM cell.

The CARRY cell is designed using GDI technique as shown in figure 7. The GDI cell is similar to inverter cell. The only difference is instead of connecting the source of the PMOS to the VDD

25

and source of NMOS to the GND, two different inputs are provided through the sources of PMOS and NMOS. For CARRY calculation again H is used as input to the Gate and A & Cin variables are connected to the Source of PMOS and NMOS respectively. The output waveform is shown in figure 9.

The logic characteristics are satisfied in PTL-GDI design. For example considering ABCin→101, as A is HIGH and B is LOW, the PMOS transistors passes the value of A (i.e. 1). In this case the pull down circuit will be inactive. So the value of H is 1. Now in the SUM circuit, as H and Cin both are HIGH, the PMOS circuit will be OFF and the pull down circuit will be ON. This will produce Logic '0' at the SUM output. On the other hand in the CARRY cell, as H is HIGH, the NMOS transistor is ON. So it will pass the value of Cin (i.e. 1) to the CARRY output.

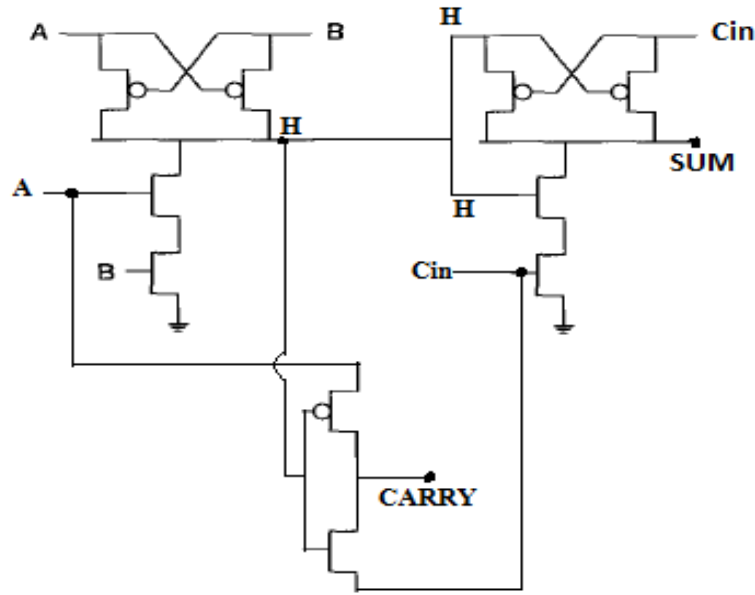

Figure 7. Proposed PTL-GDI based SUM cell

### 3.2 Proposed GDI Adder

In this design the SUM as well as CARRY cell is designed using GDI technique. It needs totally 8 transistors to implement the SUM cell and 2 transistors to design CARRY cell. Firstly H function i.e. XOR is implemented using GDI technique and then using this H function as input, the overall SUM as well as CARRY cell is been implemented. The CARRY cell design is similar to the CARRY cell of the Proposed PTL-GDI Adder. The overall circuit is shown in figure 8. The output waveform is shown in figure 10.

The logic characteristics are satisfied in Proposed GDI design too. Again considering the same example ABCin→101, as A is HIGH and B is LOW, the PMOS transistor (where B is connected as Gate input) is turned ON. This will pass the value of A (i.e. 1). Hence the value of H is 1. Now as the value of H is HIGH, the NMOS transistor of the SUM cell is turned ON. This will produce logic '0' at the SUM output. On the other hand in the CARRY cell, as H is HIGH, the NMOS transistor is ON. So it will pass the value of Cin (i.e. 1) to the CARRY output.

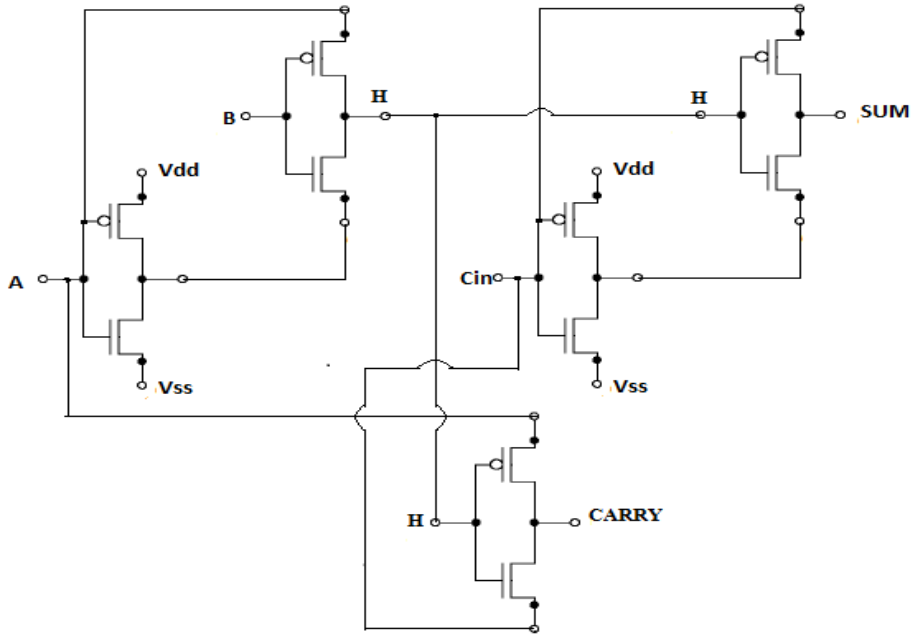

Figure 8. Proposed GDI based SUM cell

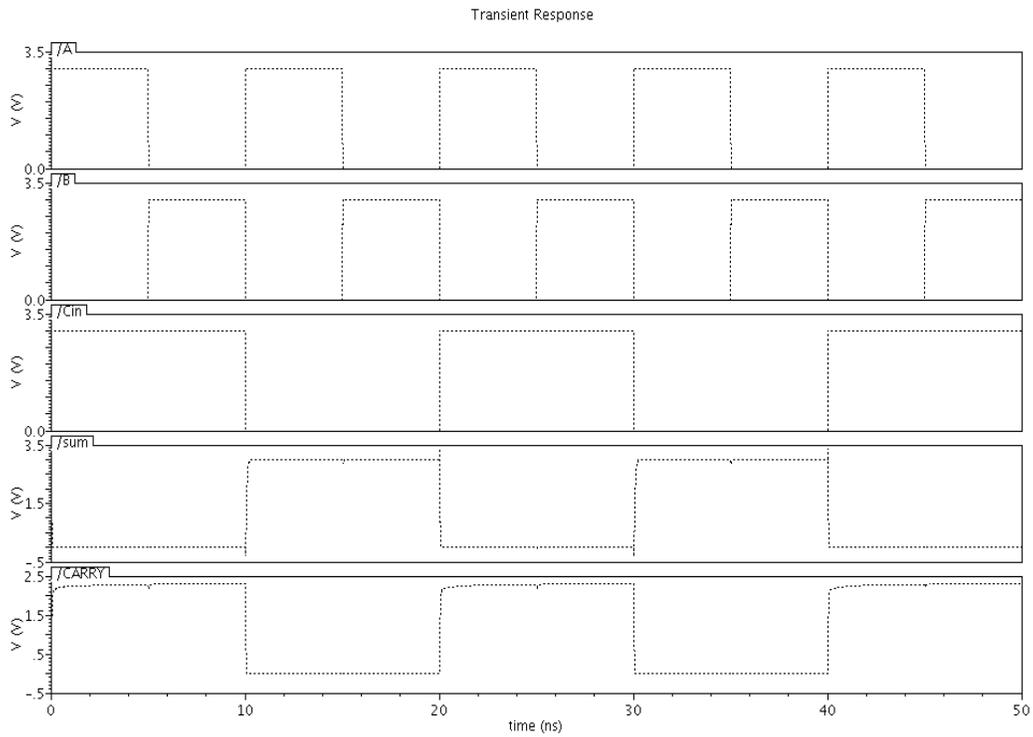

Figure 9. Output waveform of Proposed PTL-GDI Adder

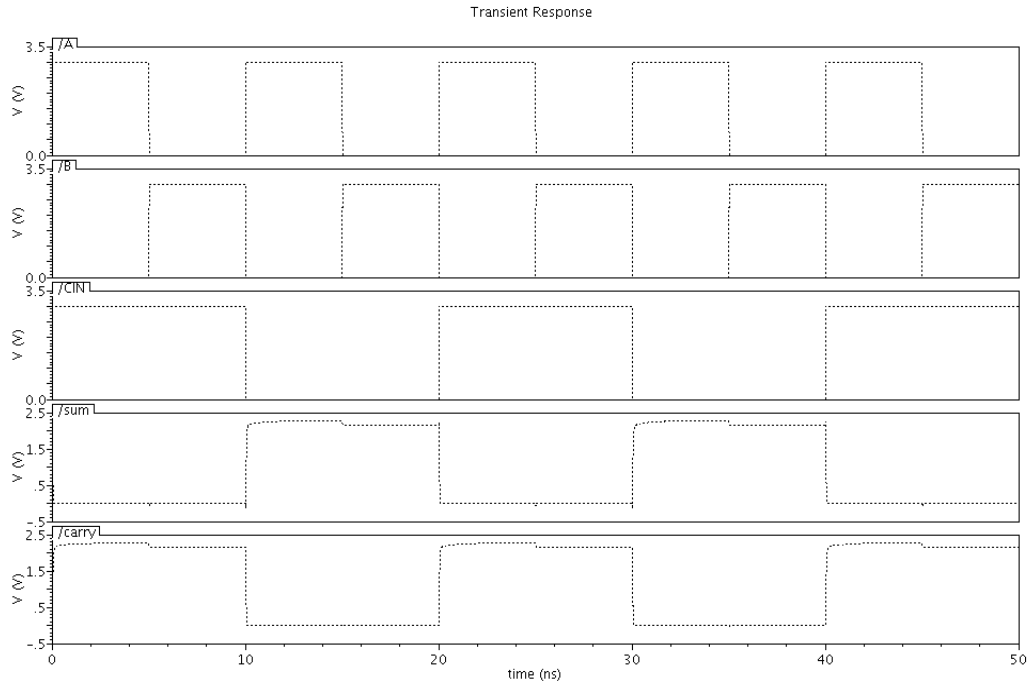

Figure 10. Output waveform of Proposed GDI Adder

## 4. SIMULATION ENVIRONMENT

All the adders are designed and simulated in Cadence Virtuoso 180nm technology. Their performances are measured in different supply voltages such as 3V, 1.8V and 0.8V at 100MHz. The delay was measured from 50% of the input voltage swing to 50% of the output voltage swing. Mainly three parameters are compared in this analysis; they are Delay, Power Consumption and Power Delay Product (PDP). In order to have a fair comparison, all the simulated circuits are prototyped at optimum transistor sizing. The transistor sizes of all the simulated circuits have been included in the figures. In the circuits, the numbers depict the width (W) of the transistors with the minimum feature size as 2µm.

### 4.1 Simulation results and discussion

By optimizing the transistor sizes of the full adders considered, it is possible to reduce the delay of all adders without significantly increasing the power consumption, and transistor sizes can be set to achieve minimum PDP. All adders were designed with minimum transistor sizes initially and then simulated. To achieve minimum PDP, an iterative process of redesigning and transistor sizing after post-layout simulations was carried out. The comparison of full adders designed to achieve minimum PDP is discussed below. In particular, three subsections refer to delay, power, and PDP respectively. In each subsection, effect of varying supply voltage is considered.

#### 4.1.1 Number of Transistor Used

The basic goal of an adder circuit is to produce correct logic characteristics with minimum number of transistors in order to produce lesser delay and optimum power consumption. As it is a basic concept that if the number of transistor decreases the delay as well as power consumption decreases. So our main motive was to reduce the number of transistor in our proposed design. Table 2 shows the number of transistors used for different Adders.

Table 2. Number of Transistor used in various adders

| NUMBER OF TRANSISTORS | | | |
|---|---|---|---|
| SL NO | ADDER | NMOS | TOTAL |
| 1 | RADHAKRISHNAN | 7 | 14 |
| 2 | AGARWAL | 19 | 32 |
| 3 | CHANG | 12 | 26 |
| 4 | GOEL | 11 | 22 |
| 5 | PROPOSED PTL-GDI | 5 | 10 |
| 6 | PROPOSED GDI | 5 | 10 |

### 4.1.2 Delay Comparison

As delay is the major issue to determine the characteristics of the design, our one main goal was to reduce the delay. The values of delay obtained for different Vdd values of 0.8V, 1.8V and 3V. To make the comparison easier, Table 3 shows the delay values of different adders at 3V, 1.8V & 0.8V. As we can see from table 2 and table 3 that when the number of transistors decreases the delay also decreases. The delay is found to be very high in the case of Agarwal as well as Goel Adder. Worst performer is Agarwal Adder at 3V Vdd. But when the supply voltage decreases to 1.8V & 0.8V the worst performer is Goel Adder. Moreover in the case of Radhakrishnan Adder, Goel Adder & Chang Adder distorted outputs are generated at 0.8V Vdd (i.e. when the supply voltage decreases beyond some threshold value these adders are found to produce unexpected output).

Table 3: Delay comparison of different adders

| DELAY(pico second) | | | | |
|---|---|---|---|---|
| SL NO | ADDER | 3V | 1.8V | 0.8V |
| 1 | RADHAKRISHNAN | 19.03 | 25.02 | 103.1 |
| 2 | AGARWAL | 59.13 | 79.74 | 448.3 |
| 3 | CHANG | 28.99 | 46.99 | 974.5 |
| 4 | GOEL | 55.68 | 92.4 | 746.6 |
| 5 | PROPOSED PTL-GDI | 17.13 | 23.29 | 93.69 |
| 6 | PROPOSED GDI | 13.8 | 19.39 | 88.35 |

Now considering the proposed designs i.e. Proposed PTL-GDI Design & Proposed GDI design, the number of transistors used here is comparatively very less. On the other hand the delay is also very low at different supply voltages for both the designs. Moreover both the proposed design produces correct logic characteristics even when the supply voltage is reduced to 0.8V. Looking at the table it can be easily understood that even though the number of transistors used for the proposed design are same but proposed GDI is found to be more delay efficient.

### 4.1.3 Power Comparison

The average power dissipation is evaluated under different supply voltages. Table 4 tabulates the values at 3V, 1.8V & 0.8V. Among the conventional existing full adders, clearly CPL has the highest power dissipation. The CPL adder dissipates the most power because of its dual-rail structure and high number of internal nodes in its design. Therefore, the CPL topology should not be used if the primary target is low power dissipation. In the comparison table as we can see,

Goel Adder consumes a huge amount of power. Even though the numbers of transistor used for Agarwal Adder is maximum, its power consumption is found to be very less with respect to Goel Adder. The proposed PTL-GDI design consumes lesser power in comparison with Radhakrishnan Adder even though the numbers of transistors used is only four more in Radhakrishnan Adder. Comparing the Proposed PTL-GDI Adder with Proposed GDI Adder, the power consumption of Proposed GDI Adder is found to be more at 3V. But as the supply voltage decreases Proposed GDI Adder is found to be best performer in the comparison table.

Table 4: Power comparison of different adders

| POWER CONSUMPTION(micro watt) | | | | |
|---|---|---|---|---|
| SL NO | ADDER | 3V | 1.8V | 0.8V |
| 1 | RADHAKRISHNAN | 12.28 | 4.044 | 312.4 pw |
| 2 | AGARWAL | 186.3 | 70.7 | 14.62 |
| 3 | CHANG | 100.8 | 26.67 | 43.09 nw |
| 4 | GOEL | 876.5 | 228.7 | 12.13 |
| 5 | PROPOSED PTL-GDI | 2.779 | 1.097 | 225.3 pw |
| 6 | PROPOSED GDI | 3.19 | 1.054 | 119.6 pw |

### 4.1.4 PDP Comparison

The PDP is a quantitative measure of the efficiency of the tradeoff between power dissipation and speed, and is particularly important when low-power operation is needed. The values of PDP are evaluated under different supply voltages are tabulated in Table 5.

Table 5: PDP comparison of different adders

| POWER DELAY PRODUCT(PDP) | | | | |
|---|---|---|---|---|
| SL NO | ADDER | 3V | 1.8V | 0.8V |
| 1 | RADHAKRISHNAN | 0.233 fj | 0.101 fj | 32.2 zj |
| 2 | AGARWAL | 11.016 fj | 5.613 fj | 6.554 fj |
| 3 | CHANG | 2.922 fj | 1.253 fj | 0.042 fj |
| 4 | GOEL | 48.803 fj | 21.13 fj | 9.056 fj |
| 5 | PROPOSED PTL-GDI | 0.048 fj | 0.026 fj | 21.11 zj |
| 6 | PROPOSED GDI | 0.044 fj | 0.020 fj | 10.57 zj |

The PDP is measured in Femto joule (fj) and Zepto Joule (zj). The Goel Adder has the maximum PDP even though the numbers of transistor used in Goel Adder is lesser than Agarwal Adder. Though the Chang Adder is found to be best in the literature [1] (with respect to PDP), both Proposed PTL-GDI Adder & Proposed GDI Adder are having very less PDP in different supply voltages.

## 5. CONCLUSION

Hybrid design style gives more freedom to the designer to select different modules in a circuit depending upon the application. Using the adder categorization and hybrid design style, many full adders can be conceived. Here two novel full adders are designed using GDI design as well as PTL-GDI design style are presented in this paper that targets low PDP. The proposed hybrid full adders have better performance than most of the standard full-adder cells owing to the novels design modules proposed in this paper. It performs well with supply voltage scaling. From the comparison table it can be inferred that both the proposed designs are good performer at different supply voltage conditions. However both the designs have their own advantages. They are:

1) If the supply voltage is above the threshold voltage (for example 3V), it is suggested to use Proposed PTL-GDI Adder.
2) If the adder is to be used in a wide range of supply voltages (for example 0.8V-3V), it is suggested to use the Proposed GDI design.
3) Considering Delay into account both the designs are found to be best in different supply voltages.

## ACKNOWLEDGEMENT

The Authors acknowledge the support of the School of Electronics Engineering (SEE) of Lovely Professional University (LPU), Phagwara, Punjab (INDIA).